\documentstyle[aps,prb,twocolumn,floats,psfig]{revtex}

\draft
\begin{document}
\twocolumn[\hsize\textwidth\columnwidth\hsize\csname@twocolumnfalse%
\endcsname

\title{Suppression of static stripe formation by next-neighbor hopping}

\author{B. Normand and A. P. Kampf}

\address{Theoretische Physik III, Elektronische Korrelationen und 
Magnetismus, Institut f\"ur Physik, \\ Universit\"at Augsburg, D-86135 
Augsburg, Germany}

\date{\today}

\maketitle

\begin{abstract}

We show from real-space Hartree-Fock calculations within the extended 
Hubbard model that next-nearest neighbor ($t^{\prime}$) hopping processes 
act to suppress the formation of static charge stripes. This result is 
confirmed by investigating the evolution of charge-inhomogeneous corral 
and stripe phases with increasing $t^{\prime}$ of both signs. We propose 
that large $t^{\prime}$ values in YBCO prevent static stripe formation, 
while anomalously small $t^{\prime}$ in LSCO provides an additional reason 
for the appearance of static stripes only in these systems. 

\end{abstract}
\bigskip
]

The presence of charge-inhomogeneous and striped phases as candidate 
ground states for the cuprates continues to be one of the most contentious 
issues in high-temperature superconductivity (for a review and references 
see Ref.~\onlinecite{rnk}). Static stripes have been observed 
experimentally, but to date only in rare-earth-doped La$_{2-x}$Sr$_x$CuO$_4$ 
(LSCO) systems with the low-temperature tetragonal (LTT) structural 
distortion. Theoretical explanations for stripes fall into three categories: 
i) they are a true ground state intrinsic to short-ranged models of the 
CuO$_2$ planes, ii) they are a competing excited state stabilized by 
anisotropy, or iii) they emerge when long-ranged interactions are 
invoked (generally to frustrate phase separation). We have recently 
subscribed to the second viewpoint, motivated by the direct experimental 
connection of lattice structure with stripe formation and suppression of 
superconductivity.\cite{rnk} 

However, the interpretation of striped phases as a consequence of 
lattice-induced hopping and superexchange anisotropy in the CuO$_2$ plane 
leaves some open questions. On the qualitative level, these include the 
absence of static stripes in orthorhombic YBa$_2$Cu$_3$O$_{7-\delta}$ 
(YBCO) systems, while quantitatively they 
include the value of the relative anisotropies required to stabilize a 
stripe phase. A key issue long recognized as a source of major differences 
between cuprate compounds is the shape of the Fermi surface.\cite{rtm} 
This has been measured by angle-resolved photoemission (ARPES) for 
Bi$_2$Sr$_2$CaCu$_2$O$_{8+\delta}$ (BSCCO),\cite{rdea} LSCO,\cite{riea} 
and Nd$_{2-x}$Ce$_x$CuO$_4$ (NCCO)\cite{rka} systems, and is 
most easily modeled by extended tight-binding band structures.\cite{rtm} 
The primary influence of an extended band structure may be encapsulated 
in the single parameter $t^{\prime}$ governing the next-nearest-neighbor 
hopping on the square lattice. An investigation of $t^{\prime}$ effects 
is required in context of stripes, and is provided here within the 
real-space Hartree-Fock (RSHF) technique.\cite{rnk}  

The effects of next-neighbor hopping on stripes have been considered in 
the extended $t$-$J$ model by exact diagonalization (ED)\cite{rtgsclmd} 
and by density-matrix renormalization group (DMRG) calculations.\cite{rws}
Both sets of authors reported a suppression of horizontal (or vertical) 
stripe order with increasing $|t^{\prime}|$ of both signs, but detailed 
features of the results were not fully consistent. The authors of 
Ref.~\onlinecite{rws} also made a systematic comparison of stripe and 
pairing instabilities, finding a strong anticorrelation. A $t^{\prime}$ 
opposite in sign to $t$ was included in a more general study of the 
extended ($t$-$t^{\prime}$) Hubbard model by RSHF,\cite{rvvg} which 
confirmed the tendency towards stripe suppression. The influence of 
$t^{\prime}$ on stripes has been considered in the context of Fermi-surface 
geometry,\cite{rim} and in terms of its effects on stripe filling.\cite{rmobb} 
Finally, extended hopping integrals were used in an ED study of doped 
stripes,\cite{rmgxfd} where their role was primarily to avoid phase 
separation. We return below to a more detailed discussion of these 
results.

The extended, anisotropic Hubbard model is given by 
\begin{eqnarray}
H & = & - \sum_{i, {\eta = x,y}, \sigma } t_{\eta} ( c_{i \pm \eta 
\sigma}^{\dag} c_{i\sigma} + H. c.) + U \sum_i n_{i \uparrow} n_{i 
\downarrow} , \label{eehh} \nonumber \\ & & \;\; - \sum_{i, {\eta = x \pm y}, 
\sigma } t_{\eta}^{\prime} ( c_{i \pm \eta \sigma}^{\dag} c_{i\sigma} + 
H. c.) ,
\end{eqnarray}
where $n_{i\sigma} =  c_{i\sigma}^{\dag} c_{i\sigma}$. We retain in 
Eq.~(\ref{eehh}) the possibility of anisotropic nearest-neighbor\cite{rnk}  
and next-neighbor hopping, $t_x \ne t_y$ and $t_{x+y}^{\prime} \ne 
t_{x-y}^{\prime}$. However, the symmetry of next-neighbor terms in 
this one-band model is maintained in the LTT phase of LSCO, and for 
most of the study to follow. We apply the HF decomposition of 
the Hubbard term in Eq.~(\ref{eehh}) and seek self-consistent solutions 
for the static charge and spin configurations.

A detailed analysis of the RSHF technique was provided in a recent study 
of anisotropy effects on stripe formation.\cite{rnk} The solutions were 
characterized as functions of 
the ``intrinsic'' system parameters, by which is meant the ratio $U/t$, 
the (hole) filling $x$, the temperature $T$, and the hopping anisotropy 
$t_x \ne t_y$, and for their dependence on the ``extrinsic'' parameters 
system size and geometry, boundary conditions (BCs), and commensuration of 
$x$ with system size. As a result of this investigation, we have chosen 
two parameter sets representative of inhomogeneous spin and charge 
configurations whose evolution we will follow on varying $t^{\prime}$ in 
the extended Hubbard model. Hereafter we take the energy unit to be $t = 1$. 

\begin{figure}[t!]
\mbox{\psfig{figure=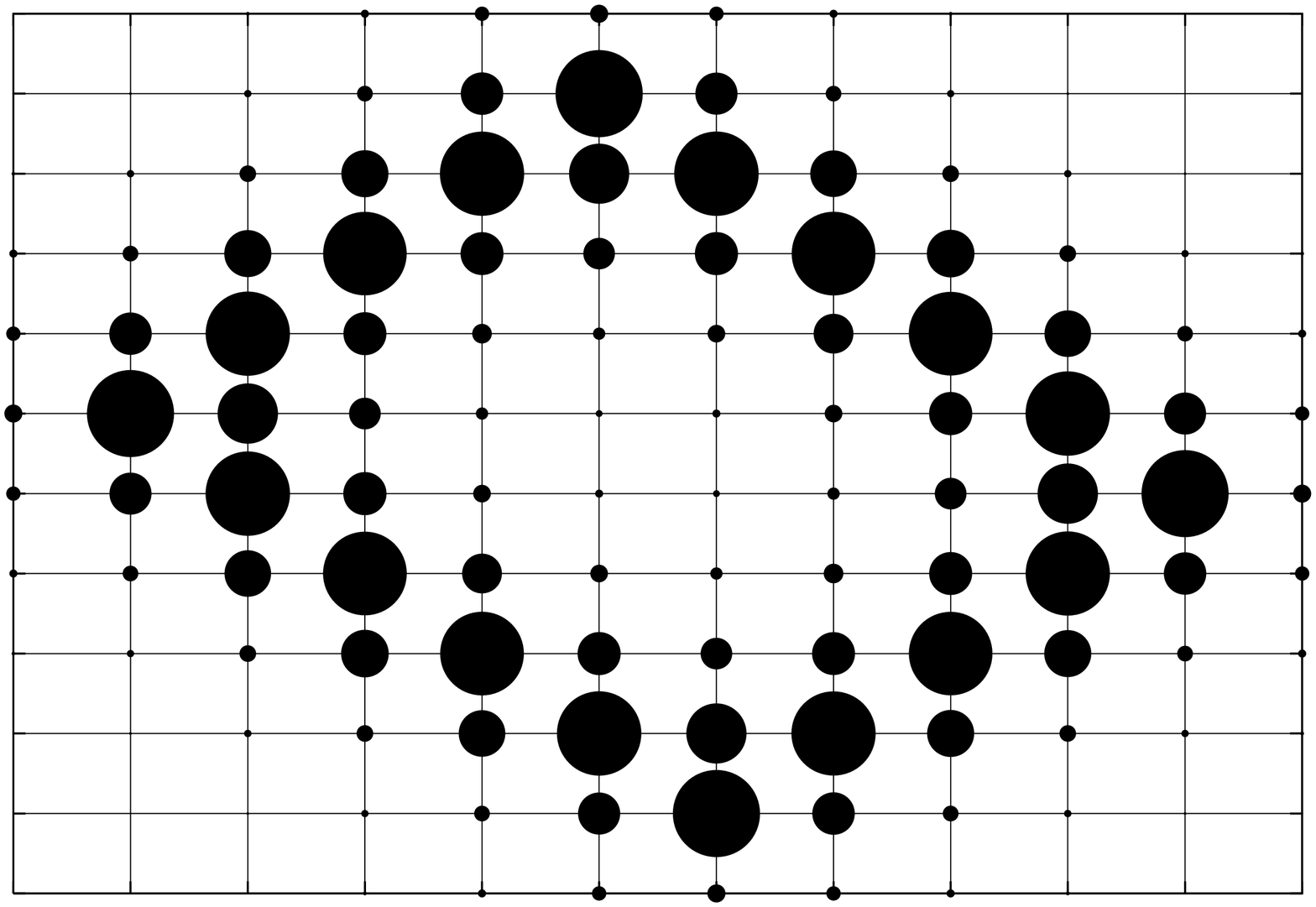,height=3.0cm,angle=270}}
\mbox{\hspace{-0.5cm}\psfig{figure=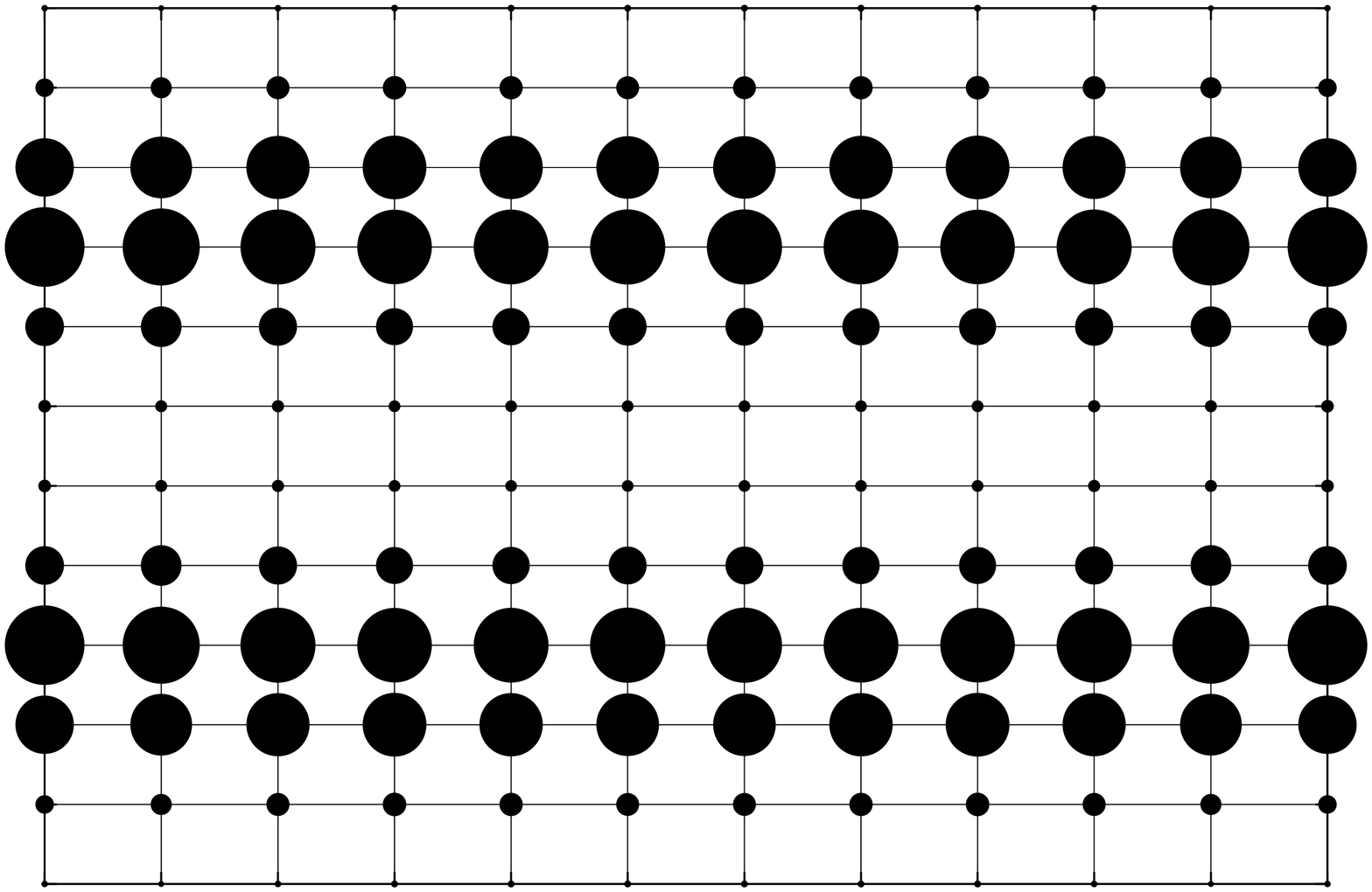,height=3.0cm,angle=270}}
\smallskip
{\centerline{(a) \qquad\qquad\qquad\qquad\qquad$\;\;$ (b)}}
\vspace{0.1cm}
\caption{Ground-state charge distributions for Hubbard model on a 
$12\times12$ cluster with $U = 5$,  $t^{\prime} = 0$, and open BCs. In (a) 
$x = 1/8$ and $t_x = t_y = 1$, while in (b) $x = 1/6$, $t_x = 0.9$ 
and $t_y = 1.1$. In these figures the hole density is scaled by radius, and 
the largest circles correspond to $\langle n_i \rangle = 0.644$, or 35.6\% 
hole doping of the site. }
\end{figure}

We focus on 12$\times$12 systems with open BCs, and take $U/t = 5$ to give 
line-like charge structures. The cases we consider are: I -- isotropic 
hopping, $t_x = t_y = 1$, with hole doping $x$ = 1/8, where the ground state 
is a corral, or closed loop of diagonal, antiphase domain walls; II -- 
anisotropic hopping, $t_x = 0.9$, $t_y = 1.1$, with hole doping $x$ = 1/6, 
where the ground state consists of uniform, filled, antiphase stripes. 
The charge configurations with $t^{\prime} = 0$ are reproduced\cite{rnk}
in Figs.~1(a) and (b) respectively for cases I and II. The site- or 
bond-centered nature of the charge structures is not important for studying 
the influence of $t^{\prime}$. With regard to stability against small 
changes in $t^{\prime}$, both corral and stripe solutions are found to have 
only minor, quantitative alterations when $|t^{\prime}| \le 0.05$ for 
both signs. 

\begin{figure}[t!]
\mbox{\psfig{figure=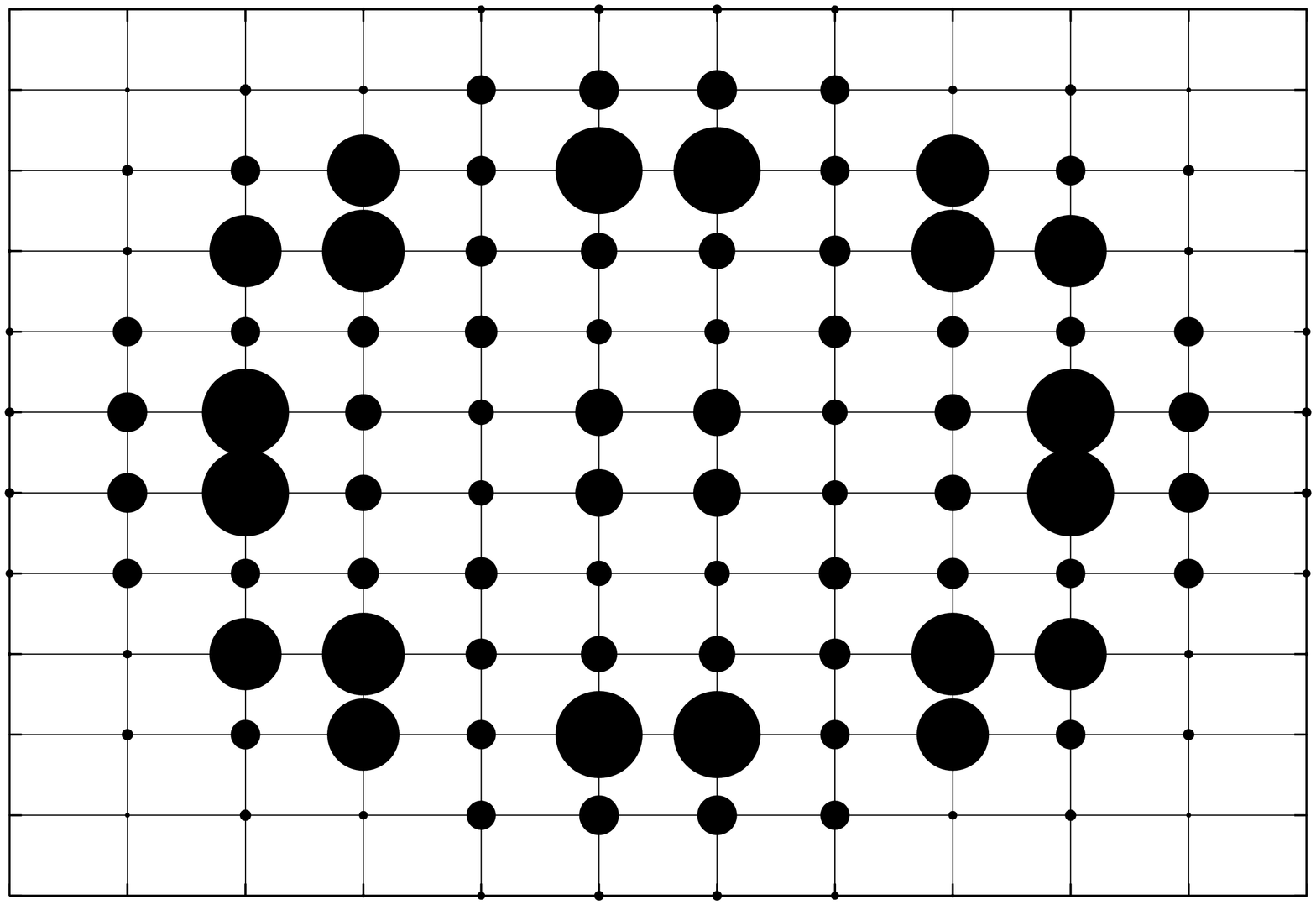,height=3.0cm,angle=270}}
\mbox{\hspace{-0.5cm}\psfig{figure=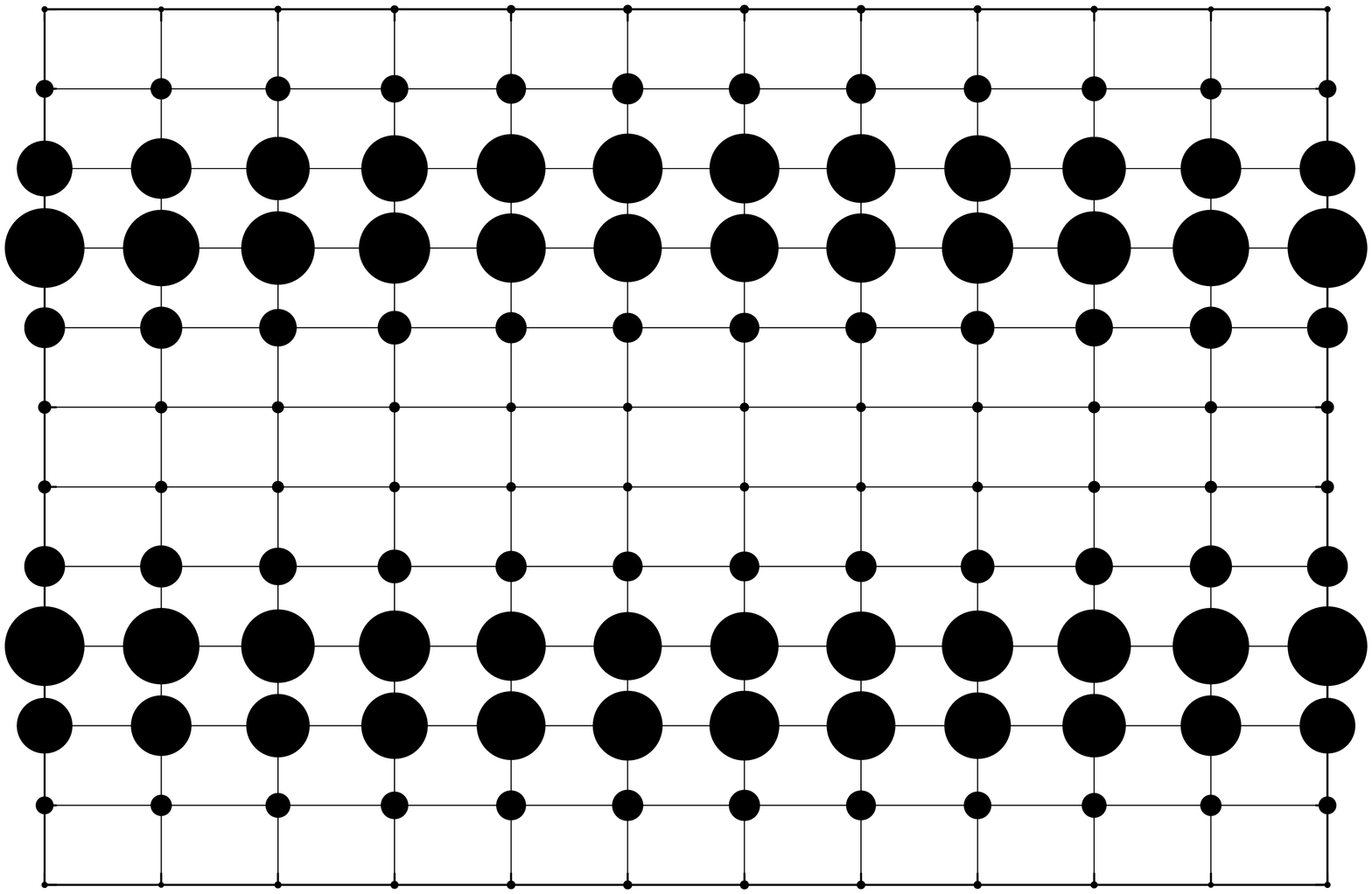,height=3.0cm,angle=270}}
\smallskip
{\centerline{(a) \qquad\qquad\qquad\qquad\qquad$\;\;$ (d)}}
\mbox{\psfig{figure=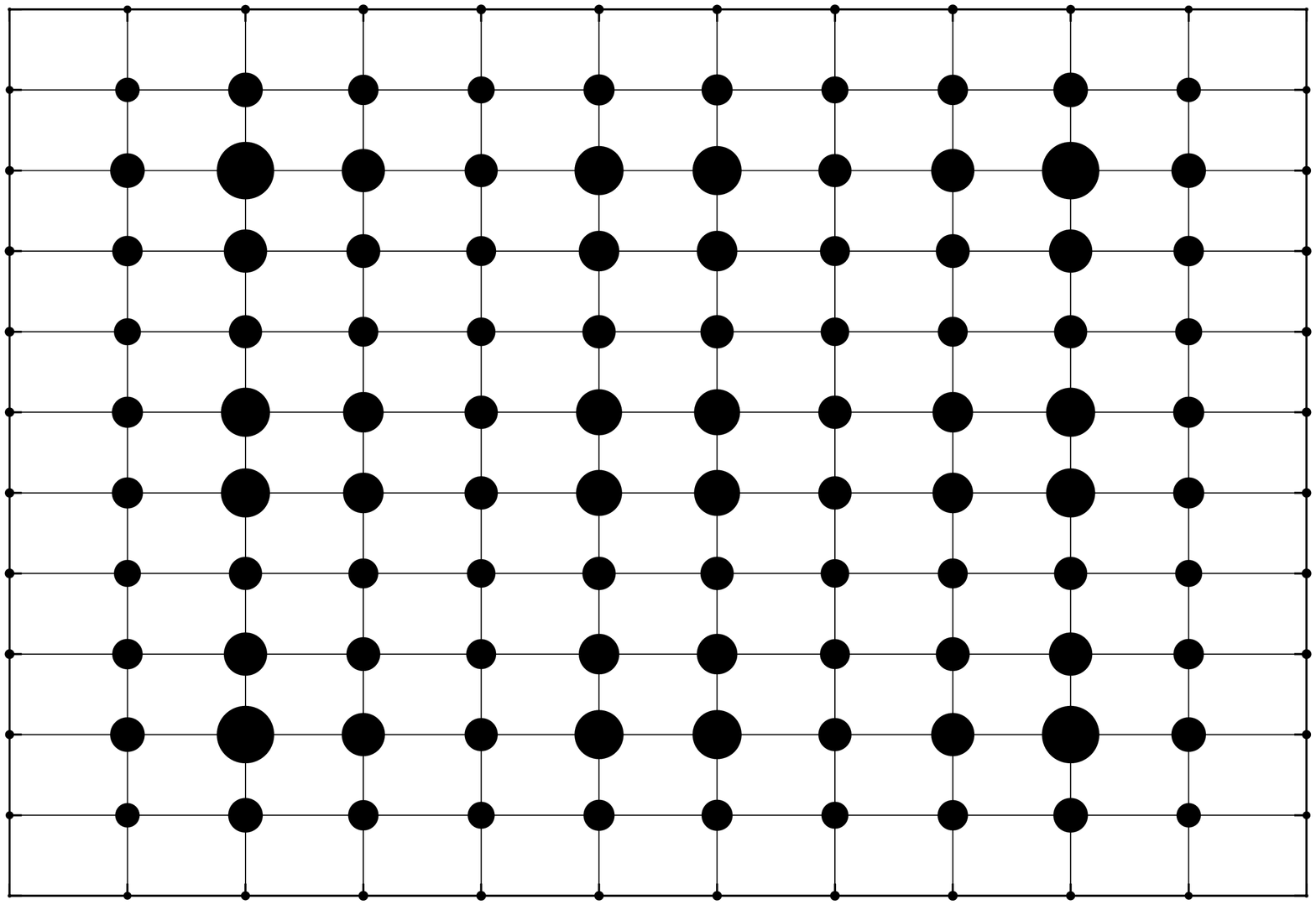,height=3.0cm,angle=270}}
\mbox{\hspace{-0.5cm}\psfig{figure=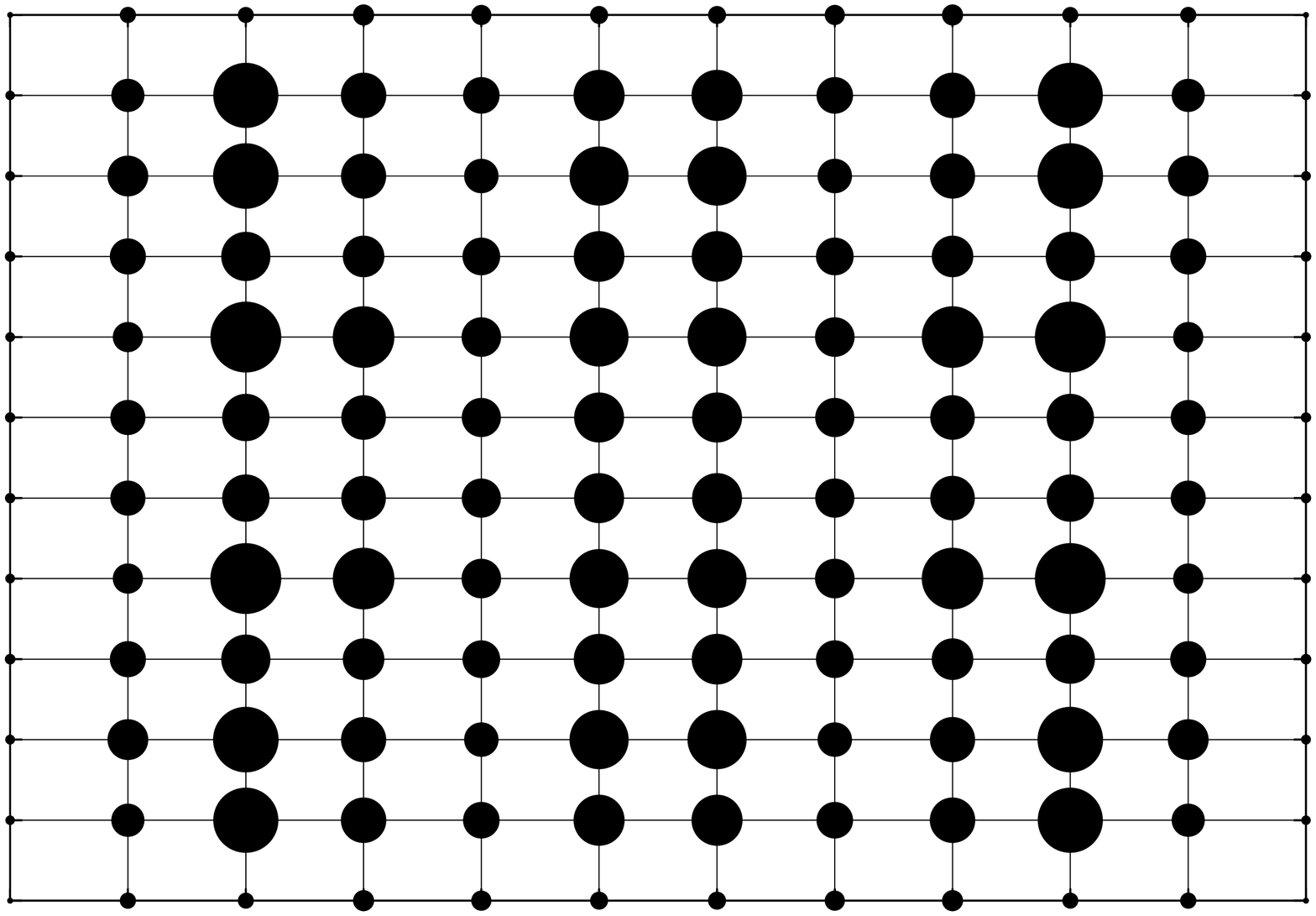,height=3.0cm,angle=270}}
\smallskip
{\centerline{(b) \qquad\qquad\qquad\qquad\qquad$\;\;$ (e)}}
\mbox{\psfig{figure=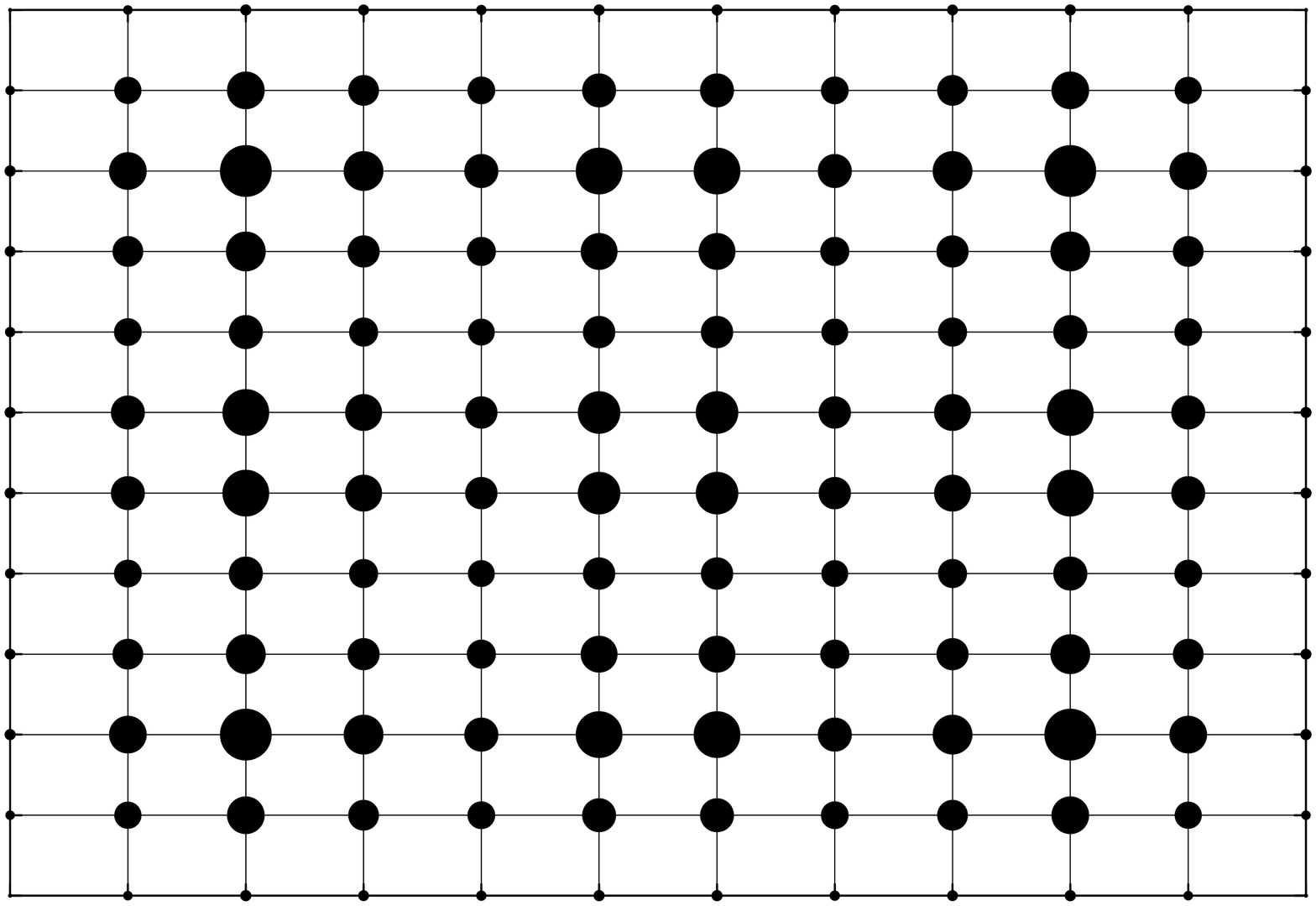,height=3.0cm,angle=270}}
\mbox{\hspace{-0.5cm}\psfig{figure=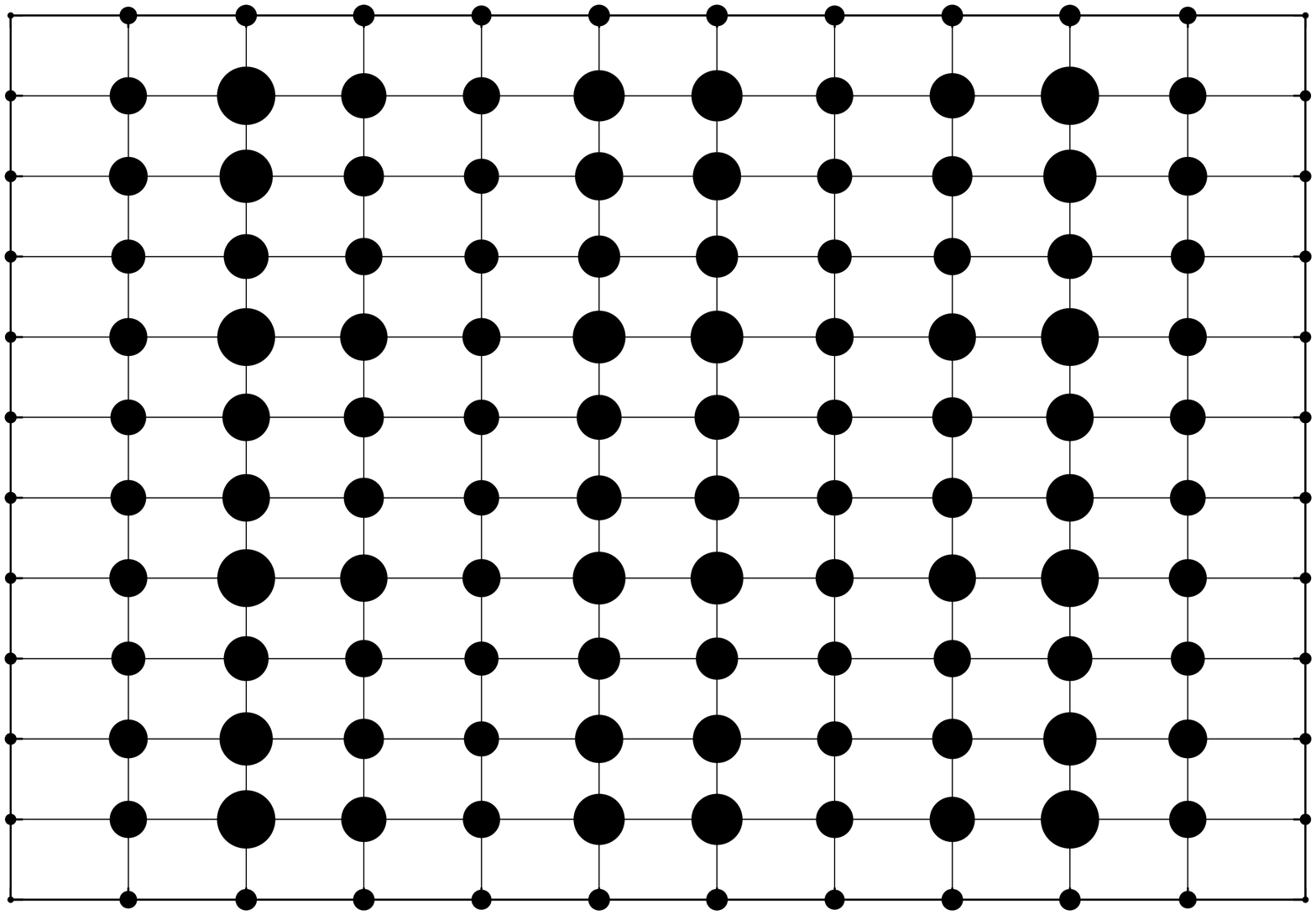,height=3.0cm,angle=270}}
{\centerline{(c) \qquad\qquad\qquad\qquad\qquad$\;\;$ (f)}}
\vspace{0.1cm}
\caption{Ground-state charge distributions for Hubbard model on a 
$12\times12$ cluster with $U = 5$ and open BCs. In (a)-(c) $x = 1/8$ and 
$t_x = t_y = 1$, while in (d)-(f) $x = 1/6$, $t_x = 0.9$ and $t_y = 1.1$. 
$t^{\prime} = 0.1$ in (a) and (d), $t^{\prime} = 0.2$ in (b) and (e), and 
$t^{\prime} = 0.3$ in (c) and (f). Charge scale as in Fig.~1. }
\end{figure}

Fig.~2 illustrates the effects of increasing a next-neighbor hopping 
$t^{\prime}$ with the same sign as $t$. This results in a change of shape 
of the Fermi surface, which remains closed around the $\Gamma$ point but 
expands in the directions $(k,\pm k)$ while contracting in the $(\pm k,0)$ 
and $(0,\pm k)$ directions. For the corral solution we see at 
$t^{\prime} = 0.1$ [Fig.~2(a)] a trend towards breaking of the domain line 
into small clusters, followed before $t^{\prime} = 0.2$ [Fig.~2(b)] by a 
complete evaporation of the inhomogeneous charge structure. The solution 
here and for higher $|t^{\prime}|$ is a charge-uniform, ``metallic'' phase. 
For the stripe solution the behavior is similar, in that the stable stripe 
phase is lost in the range $0.1 < t^{\prime} < 0.2$ in favor of the 
uniform phase, which then persists to high values of $t^{\prime}$. 
The spin configuration (not shown) in the uniform phase is found to be 
commensurate antiferromagnetism, although with an ordered moment suppressed 
in comparison to the antiferromagnetic (AF) regions of the corral and 
stripe solutions at $t^{\prime} = 0$. 

\begin{figure}[t!]
\mbox{\psfig{figure=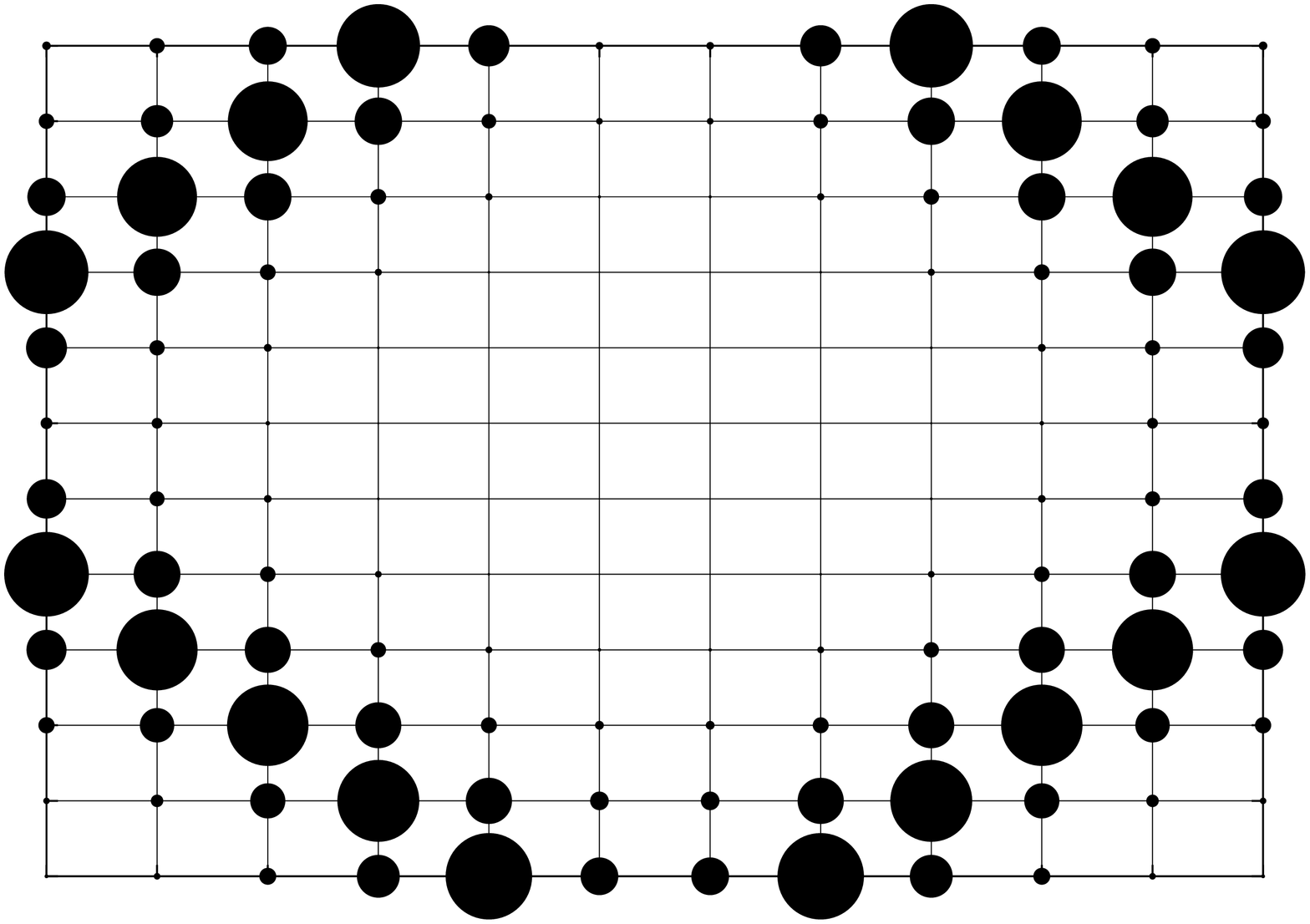,height=3.0cm,angle=270}}
\mbox{\hspace{-0.5cm}\psfig{figure=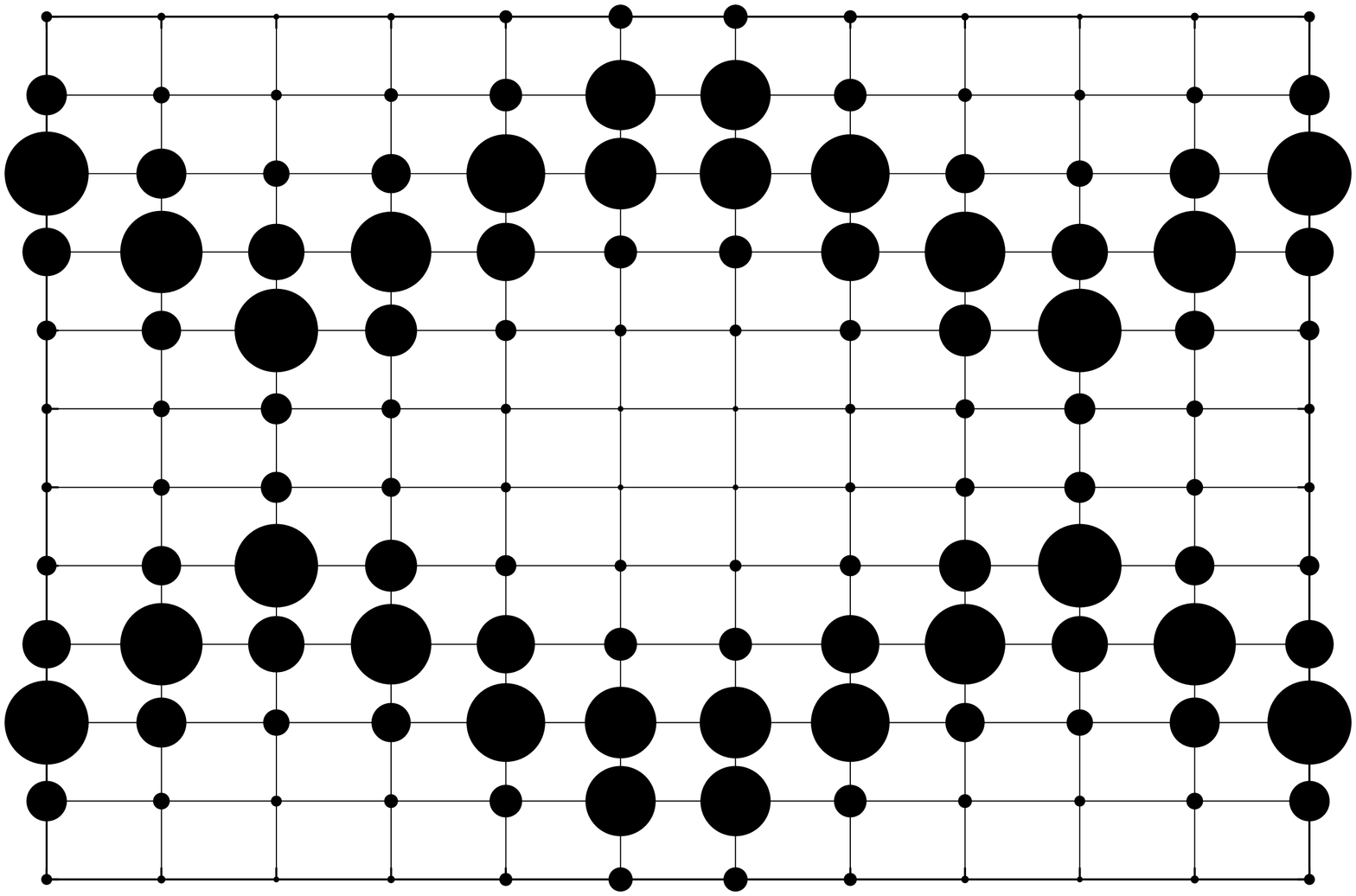,height=3.0cm,angle=270}}
\smallskip
{\centerline{(a) \qquad\qquad\qquad\qquad\qquad$\;\;$ (d)}}
\mbox{\psfig{figure=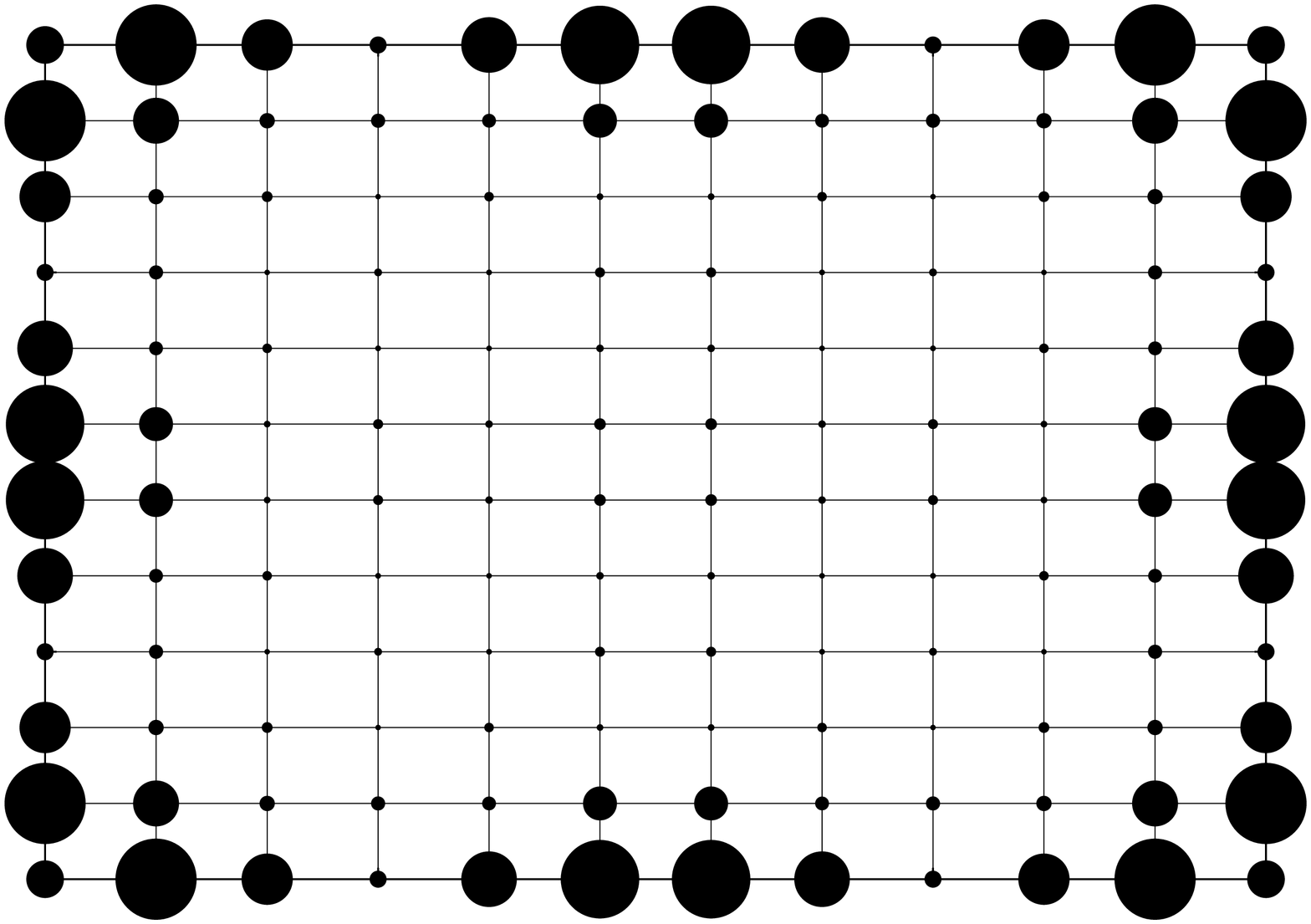,height=3.0cm,angle=270}}
\mbox{\hspace{-0.5cm}\psfig{figure=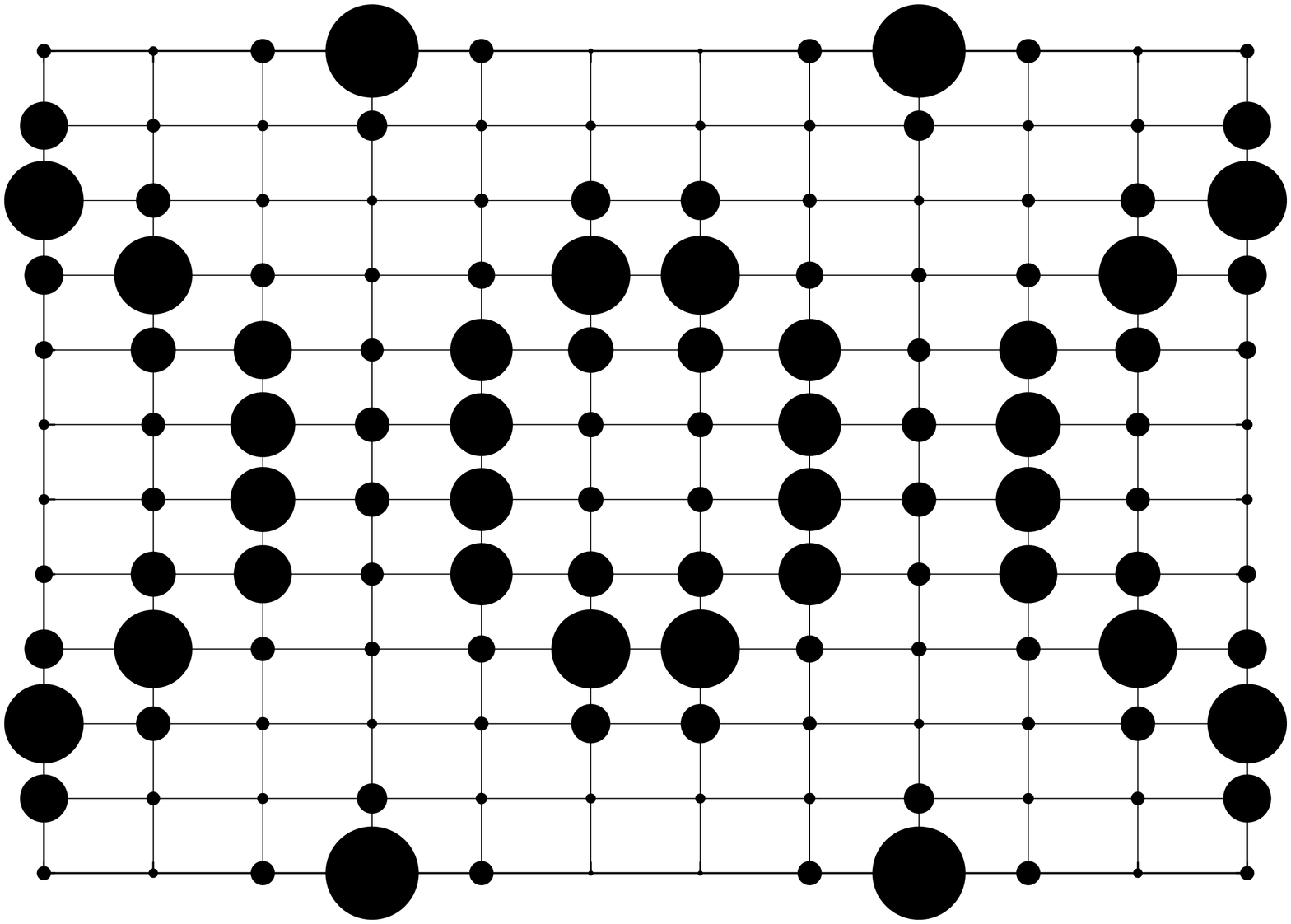,height=3.0cm,angle=270}}
\smallskip
{\centerline{(b) \qquad\qquad\qquad\qquad\qquad$\;\;$ (e)}}
\mbox{\psfig{figure=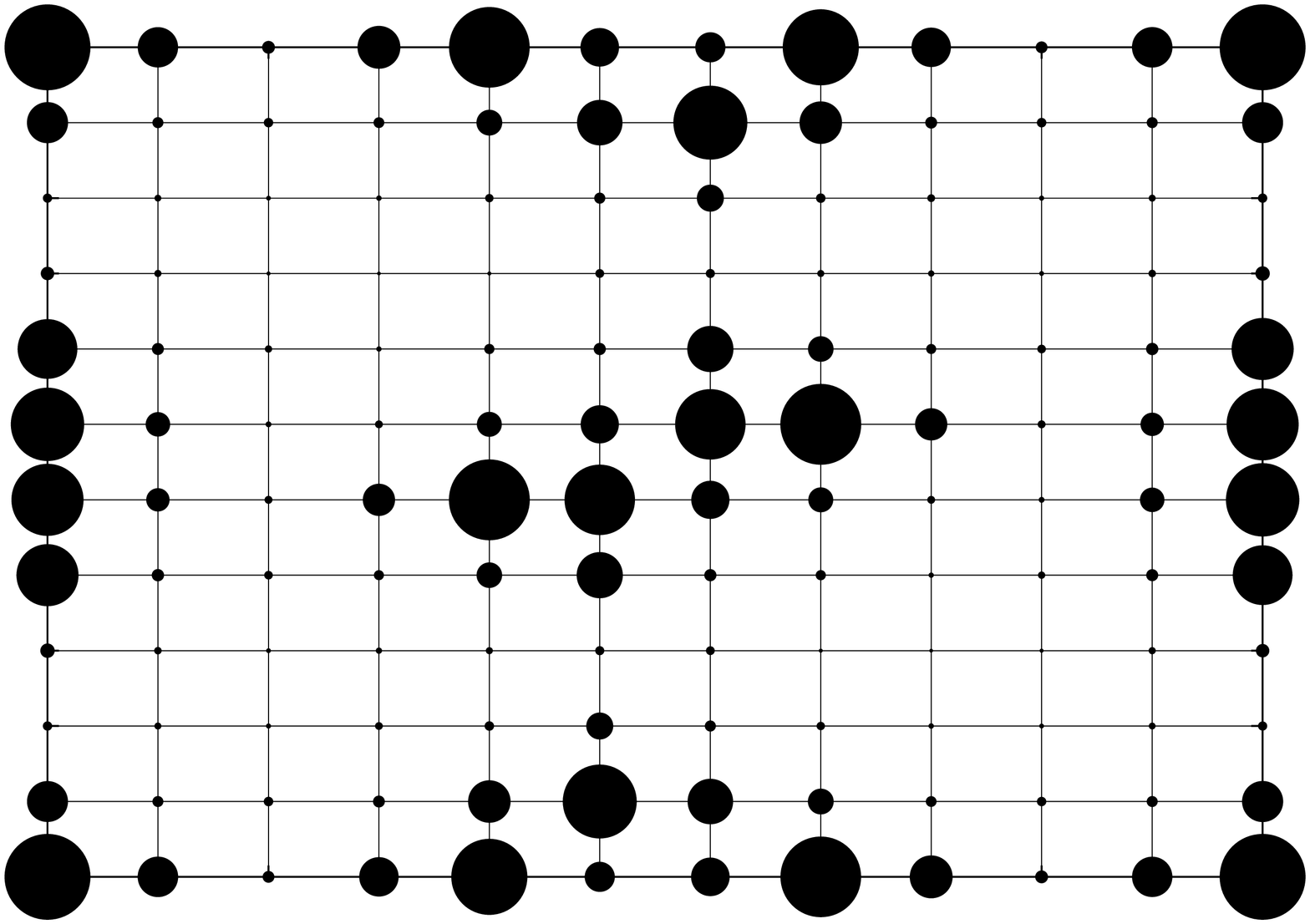,height=3.0cm,angle=270}}
\mbox{\hspace{-0.5cm}\psfig{figure=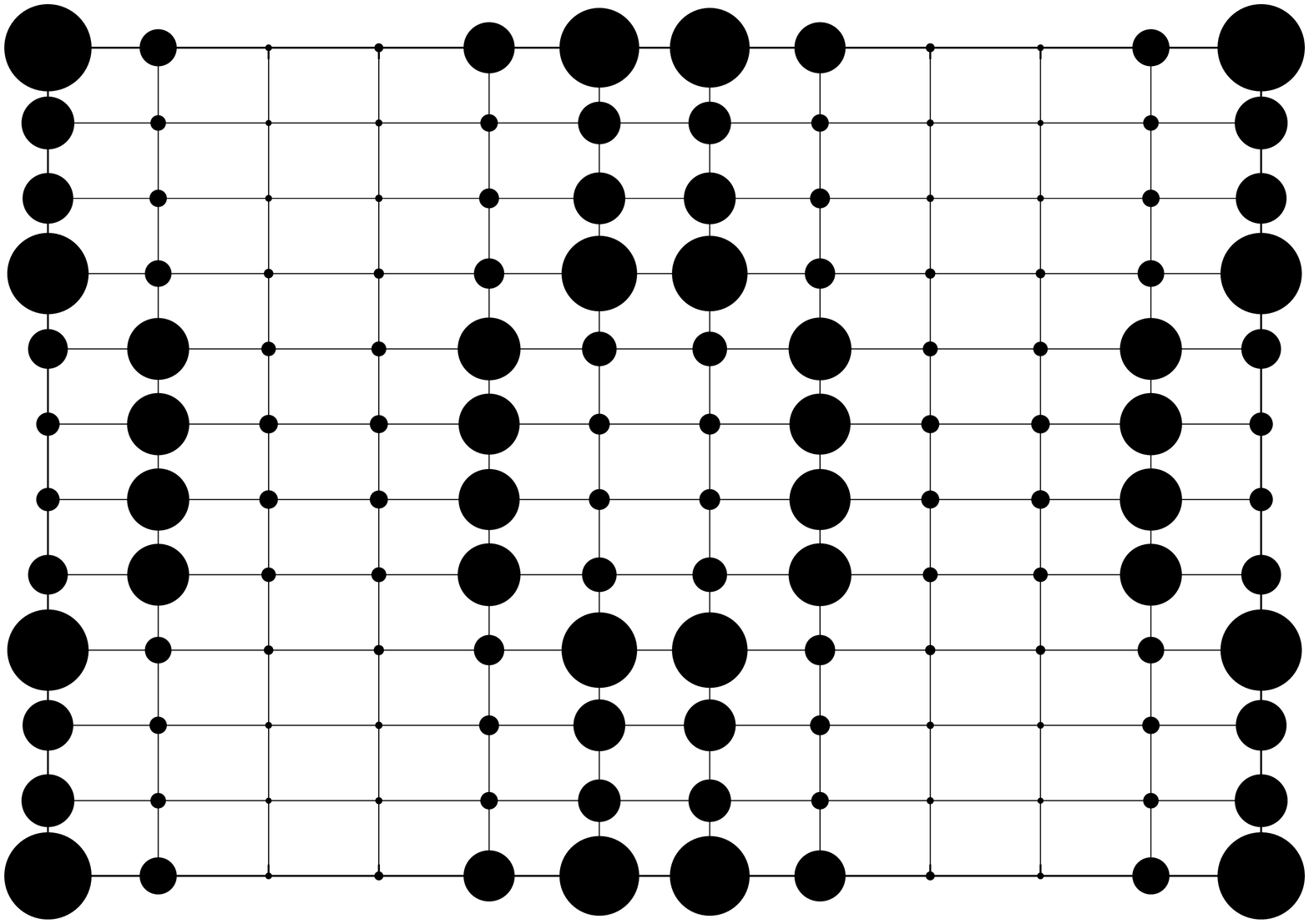,height=3.0cm,angle=270}}
\smallskip
{\centerline{(c) \qquad\qquad\qquad\qquad\qquad$\;\;$ (f)}}
\vspace{0.1cm}
\caption{Ground-state charge distributions for Hubbard model on a 
$12\times12$ cluster with $U = 5$ and open BCs. In (a)-(c) $x = 1/8$ and 
$t_x = t_y = 1$, while in (d)-(f) $x = 1/6$, $t_x = 0.9$ and $t_y = 1.1$. 
$t^{\prime} = - 0.1$ in (a) and (d), $t^{\prime} = - 0.2$ in (b) and (e), 
and $t^{\prime} = - 0.3$ in (c) and (f). Charge scale as in Fig.~1. }
\end{figure}

This result has a straightforward interpretation in terms of the metallic 
state maximizing the hopping kinetic energy. Positive $t^{\prime}$ terms 
cause direct frustration of local charge and spin alignment by competing 
with the $t$ term, resulting in a uniform distribution. One may also 
consider a band picture, where for this sign of $t^{\prime}$ the bandwidth 
of the lower Hubbard band in the spin-density wave (SDW) state at 
half-filling is increased, suggesting that doped hole-like carriers favor 
delocalization.\cite{rvvg} Our result and interpretation are fully consistent 
with DMRG studies of the $t$-$J$ model,\cite{rws} which for $t^{\prime} > 0$ 
show a systematic loss of stripe charge order. These also find an enhancement 
of hole pairing, which is further consistent with our result because 
superconductivity would be expected as the leading low-temperature
instability of the isotropic, metallic state. ED calculations\cite{rtgsclmd} 
in this regime differ in that they report enhanced stripe order for $0 < 
t^{\prime} < 0.2$, albeit for diagonal stripes at smaller (physical) values 
of $J/t$. The authors interpret this result in terms of the explicit 
next-neighbor hopping required to cancel the intra-sublattice hopping 
generated implicitly by the $t$ term, and in accord with the present 
findings expect that carrier localization and stripe-formation are 
maximized when the sum of these two contributions to $t^{\prime}$ is 
minimized. A DMRG study\cite{rtgsclmd} which tends to support this result 
remains in minor contradiction to Ref.~\onlinecite{rws}, and the resolution 
is presumably to be found in commensuration effects on the differing system 
sizes used. 

Fig.~3 illustrates the effects of increasing $|t^{\prime}|$ in the physical 
regime for the cuprates, where $t^{\prime}$ has the opposite sign to $t$. 
This results in the change from a closed to an ``open'' Fermi surface around 
the $\Gamma$ point. For the 
corral solution we see at $t^{\prime} = - 0.1$ and $- 0.2$ [Figs.~3(a,b)] 
a reinforcement of undisturbed AF order in the center of the system, and 
retention of the diagonal domain walls, until a more complex behavior 
sets in at $t^{\prime} = - 0.3$ [Fig.~3(c)]. For the stripe solution there 
is a definite crossover from horizontal stripes at small $|t^{\prime}|$ to 
diagonal stripes at $t^{\prime} = - 0.1$ and $- 0.2$ [Figs.~3(d,e)], again 
preceding a less clear structure at $t^{\prime} = - 0.3$ [Fig.~3(f)]. For 
both parameter sets, the high-$t^{\prime}$ configurations shown in Fig.~4 
are rather complex in shape, only weakly inhomogeneous in charge structure 
and weakly but commensurately AF in spin structure.

For $t^{\prime} / t < 0$, the narrowing of the lower SDW Hubbard band 
suggests reduced carrier mobility and thus an enhanced tendency toward 
charge inhomogeneity. In a local hopping picture, all next-neighbor processes 
cost kinetic energy, and so are suppressed\cite{rnk} by i) aligning spins 
ferromagnetically on diagonal bonds, promoting AF order, and ii) aligning 
holes on diagonal bonds, promoting diagonal domain walls. Both of these 
tendencies are clear in Fig.~3. At higher $t^{\prime}$ this trend cannot be 
sustained, as the kinetic energy cost of failure to delocalize is too great, 
and for $- 0.4 < t^{\prime} < - 0.3$ the kinetic energy associated with 
$t^{\prime}$ becomes negative despite the sign, as shown in Fig.~5. This is 
possible by a change in the coefficients of the HF wave function such that on 
average the sign of the diagonal overlap is negative. While such an alteration 
is not readily visualized in terms of charge configurations, its consequences 
are the weakly inhomogeneous structures in Fig.~4. The tendency to carrier 
delocalization also for this sign of $t^{\prime}$ is in accord with the 
results of Ref.~\onlinecite{rmobb} for the hole filling of an isolated stripe. 

\begin{figure}[t!]
\mbox{\psfig{figure=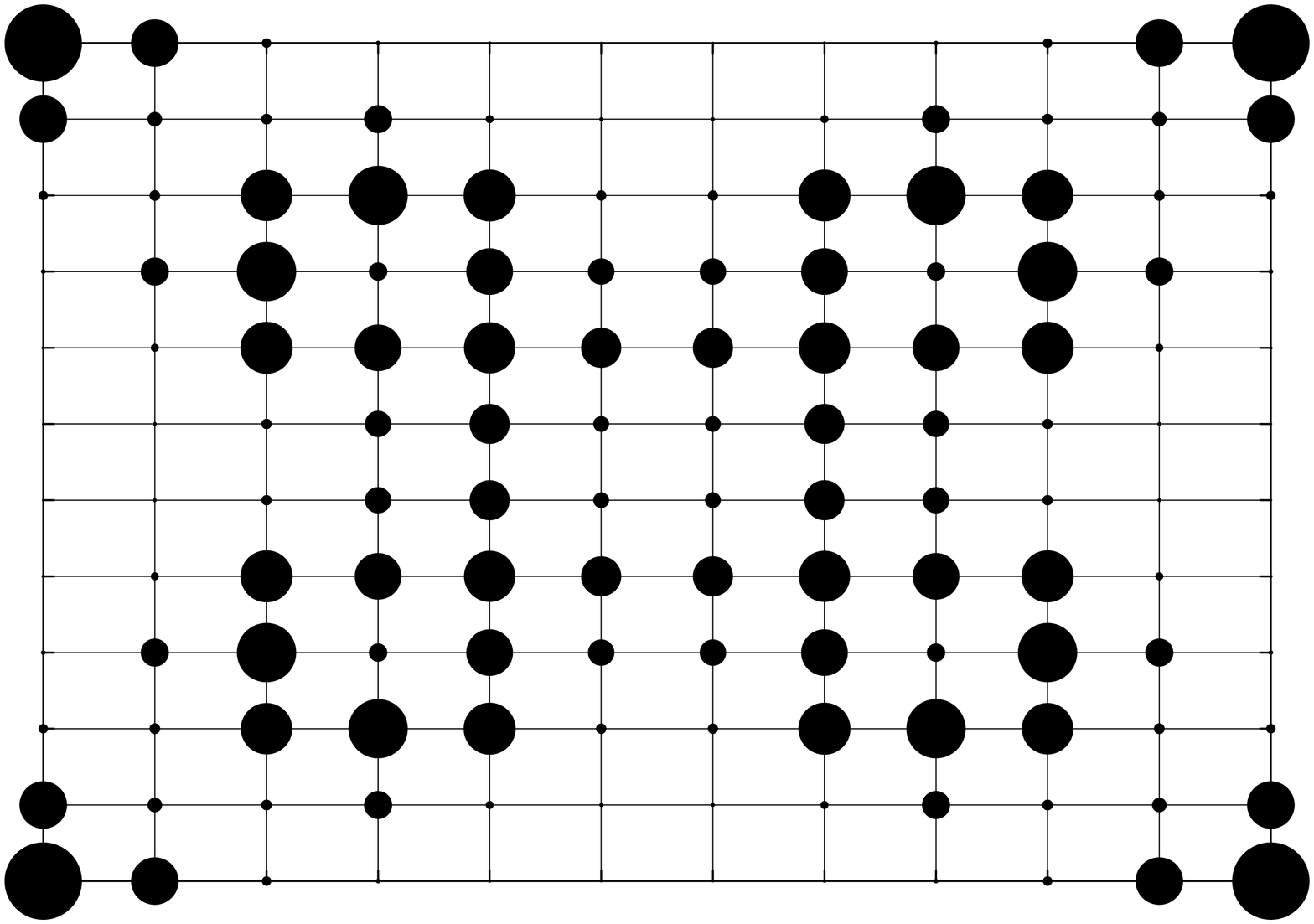,height=3.0cm,angle=270}}
\mbox{\hspace{-0.5cm}\psfig{figure=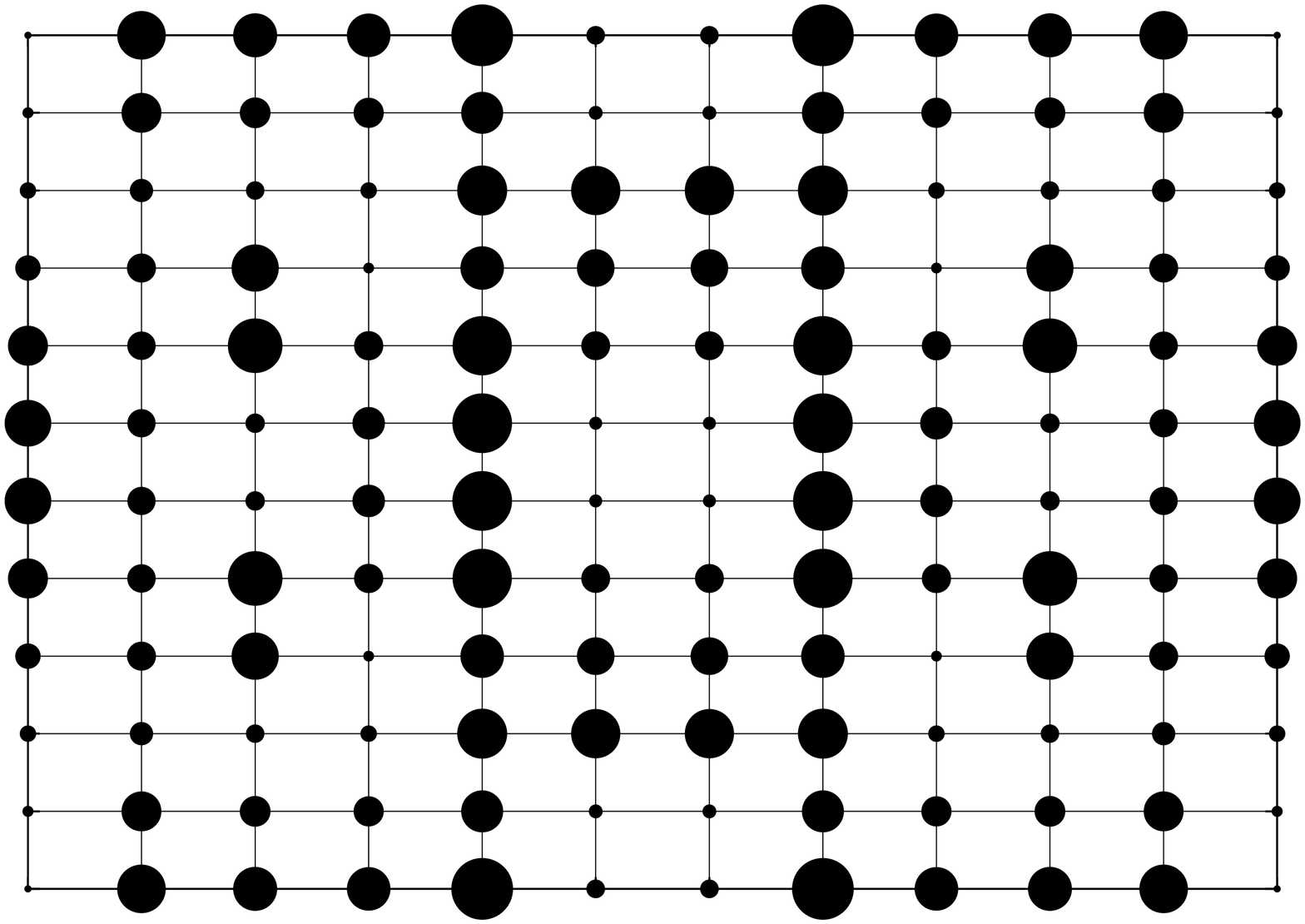,height=3.0cm,angle=270}}
\smallskip
{\centerline{(a) \qquad\qquad\qquad\qquad\qquad$\;\;$ (b)}}
\caption{Ground-state charge distributions for Hubbard model on a 
$12\times12$ cluster with $U = 5$, $t^{\prime} = 0.6$, and open BCs. In 
(a) $x = 1/8$ and $t_x = t_y = 1$, while in (b) $x = 1/6$, $t_x = 0.9$ 
and $t_y = 1.1$. Charge scale as in Fig.~1. }
\end{figure}

In the DMRG calculations,\cite{rtgsclmd,rws} increasing $|t^{\prime}|$ 
up to 0.3 introduces a doubling of the charge periodicity. From the above 
considerations we may offer a change from horizontal to diagonal stripes as 
a possible consistent explanation of this result, given the small number 
of holes involved. The observed 
suppression of pairing\cite{rws} is certainly consistent with the expectation 
of charge localization accompanying the formation of static, inhomogeneous 
charge structures. In the ED results,\cite{rtgsclmd} both vertical and 
diagonal stripe correlations are strongly suppressed by $t^{\prime}$, 
whereas pairing correlations remain appreciable. Other RSHF studies of 
the extended Hubbard model\cite{rvvg} confirm a significantly reduced 
tendency to stripe formation in this parameter regime. 

All of the above results are for hole-doped systems. In the electron-doped 
regime, we find by RSHF an exact reproduction 
of the charge configurations in Figs.~1-4, with the important proviso that 
the sign of $t^{\prime}$ be reversed. This symmetry emerges from the band 
and local physics arguments invoked above, and can be seen in Fermi-surface 
shapes, but most importantly can be proven rigorously from the invariance 
of the Hamiltonian [Eq.~(1)] under particle-hole transformation combined 
with sign reversal $c_i \rightarrow - c_i$ on one sublattice. 

In summary, next-neighbor hopping terms $t^{\prime}$ lead to a rapid 
suppression of the $t^{\prime} = 0$ inhomogeneous charge structures in 
the Hubbard model. For hole- (electron-) doped systems the primary influence 
of positive (negative) $t^{\prime}$ (Fig.~2) on stripes lies in its 
delocalization effect, while that of negative (positive) $t^{\prime}$ 
(Figs.~3,4) is destruction of the spin registry and antiphase domain wall 
nature. This latter result suggests the importance of AF order in 
stabilizing static horizontal stripes.\cite{rcn}

\begin{figure}[t!]
\centerline{\psfig{figure=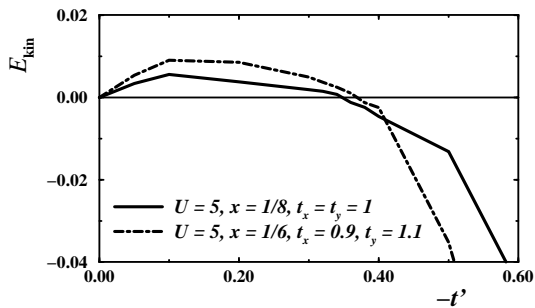,width=7.5cm,angle=270}}
\medskip
\caption{ Average kinetic energy $E_{\rm kin}$ per diagonal bond as a 
function of next-neighbor hopping $t^{\prime}$ opposite in sign to $t$.}
\end{figure}

In cuprates ($t^{\prime}/t < 0$), the effective $t^{\prime}$ is intrinsically 
smaller for LSCO and related monolayer systems than for YBCO, BSCCO and the 
related mono-, bi-, and trilayer systems. This result may be obtained by 
simple fitting of tight-banding models to the Fermi surfaces measured by 
ARPES\cite{rtm} and is also confirmed by reduction (downfolding) of band 
structure calculations.\cite{raljp} In addition to the number of layers, 
the apical oxygen atoms are seen to be an important factor contributing to 
the difference. Because deviations of the Cu--O--Cu bond angle 
from 180$^0$ reduce orbital overlap, the LTT and low-temperature orthorhombic 
(LTO) distortions of the LSCO structure also act to reduce $|t^{\prime}|$ in 
comparison with unbuckled cuprate layers. At the tight-binding level, when 
$t$ and $t^{\prime}$ are the only parameters in the extended hopping band 
structure, $t^{\prime}$ values of order $- 0.5 t$ are required to model the 
open, YBCO-like Fermi surface, while the closed LSCO surface is reproduced 
by values of $|t^{\prime}| < 0.2$. 

We have considered previously the possible key role of the LTT distortion, 
which is the ground-state lattice structure only in Nd- and Eu-doped LSCO 
compounds.\cite{rnk} Only the LTT distortion produces anisotropic $t$ terms, 
and to the best of our knowledge there remain no reports of static stripes 
in any non-LTT LSCO structures, or in any other cuprates. A natural question 
then arises concerning YBCO, where the presence of ${\hat b}$-axis chains 
leads to a planar structural 
anisotropy. Our results offer an additional explanation for the absence of 
static stripes in YBCO: the large effective $t^{\prime}$ values [Fig.~4] 
suppress the formation of charge-inhomogeneous structures which are 
candidate ground states at smaller $t^{\prime}$. Further, $t^{\prime}$ may 
be expected to be an important factor contributing to the absence of static 
stripes in the majority of hole-doped cuprate materials, where the Fermi 
surfaces of the CuO$_2$ planes are open. For the electron-doped systems, we 
have shown these to be in the regime where stripe formation is suppressed by 
any value of $t^{\prime}$, which for these dopings favors a uniform, 
metallic phase.

We comment briefly that the LTO structure of the CuO$_2$ planes in LSCO 
and BSCCO has an asymmetrical $t^{\prime}$. In RSHF, and for $t^{\prime}$ 
opposite in sign to $t$, a sufficiently large anisotropy $t_{x+y}^{\prime} 
> t_{x-y}^{\prime}$ yields diagonal stripes in the direction ${\hat x}+{\hat 
y}$. However, the required anisotropies are physically unjustified, and no 
effect of this nature is expected in the real material. 

In conclusion, we have studied the effects of a next-neighbor hopping term 
within the extended Hubbard model on the formation of static stripes in 
cuprate systems. In a RSHF treatment, $t^{\prime}$ terms of both signs 
cause a rapid suppression of horizontal stripe formation. When $t^{\prime}$ 
has the physical sign for cuprates (opposite to $t$), we find for hole-doped 
systems a window in which diagonal stripes are favored, and then a broad 
regime at large $t^{\prime}$ where the compromise charge configuration is 
only weakly inhomogeneous. For the opposite sign of $t^{\prime}$, a 
homogeneous, metallic phase is preferred. These properties are exactly 
reversed in electron-doped systems. 

In contradiction to observations of superconductivity, the ``physical'' case 
is dominated by charge localization and suppression of hole pairing, whereas 
competition of inhomogeneous charge structures with the superconducting 
instability is reduced for $t^{\prime} / t > 0$. However, these results are 
consistent with previous studies of extended Hubbard and $t$-$J$ models. We 
propose that the weak $t^{\prime}$ terms characteristic of the LSCO system 
in comparison with other high-$T_c$ superconductors is an important 
contributing factor to the appearance of static stripes only in these 
materials, and that stripe formation in electron-doped cuprates is 
suppressed for all $t^{\prime}$.

\bigskip

We are grateful to F. Guinea and C. Morais Smith for helpful discussions. 
This work was supported by the Deutsche Forschungsgemeinschaft through SFB 
484.

\end{document}